\documentclass{iopart}
\usepackage[dvips]{graphicx}
\usepackage{iopams}
\usepackage{cite}
\begin{document}
\title{Entangling two atoms via spontaneous emission}
\author{R Tana\'s\dag and Z
  Ficek\ddag\ } 
\address{\dag\ Nonlinear Optics Division, Institute of Physics, Adam Mickiewicz
University, Pozna\'n, Poland\\
\ddag Department of Physics, School of Physical Sciences, The
University of Queensland, Brisbane, QLD 4072, Australia}

\ead{tanas@kielich.amu.edu.pl}

\begin{abstract}
We discuss the creation of entanglement between two two-level
atoms in the dissipative process of spontaneous emission. 
It is shown that spontaneous emission can lead to a transient
entanglement between the atoms even if the atoms were prepared
initially in an unentangled state. The amount of entanglement created
in the system is quantified by using two different measures:
concurrence and negativity. We find analytical formulas for the
evolution of concurrence and negativity in the system. We also find
the analytical relation between the two measures of entanglement.
The system consists of two two-level atoms which are separated by an
arbitrary distance $r_{12}$ and interact with each other via the
dipole-dipole interaction, and the antisymmetric state of the
system is included throughout, even for small inter-atomic
separations, in contrast to the small sample model. It is shown that
for sufficiently large values of the dipole-dipole interaction
initially the entanglement exhibits oscillatory behaviour with considerable
entanglement in the peaks. For longer times the amount of entanglement
is directly related to the population of the slowly decaying
antisymmetric state.
\end{abstract}

\submitto{\JOB}
\pacs{32.80.-t, 42.50.-p}
\maketitle

\section{Introduction}
Entanglement is a property of quantum systems to exhibit correlations
that cannot be accounted for classically.
Entangled states of collective quantum systems, which are
states that cannot be factorized into product states of the
subsystems, are of fundamental interest in quantum
mechanics. A number of methods for creating
entanglement have been proposed involving trapped and cooled ions or
neutral atoms~\cite{bbtk,koz,hal,flei,sm01,b2,nat1,nat3}. Of particular
interest is generation of
entangled states in two-atom systems, since they can represent two
qubits, the building blocks of the quantum gates that are
essential to implement quantum protocols in quantum information processing.
It has been shown that entangled states in a two-atom system can be
created by a continuous driving of the atoms with a coherent
or chaotic thermal field~\cite{sm01,afs,klak}, or by a pulse
excitation followed by a continuous observation of radiative
decay~\cite{phbk,b1,cabr}. Moreover, the effect of spontaneous emission
on initially prepared entangled state has also been
discussed~\cite{gy1,gy2,bash,jak}. These studies,
however, have been limited to the small sample (Dicke) model~\cite{dic}
or the situation involving noninteracting atoms strongly coupled to
a cavity mode. The difficulty of the Dicke model is that it does not
include the dipole-dipole interaction among the atoms and does not
correspond to realistic experimental situations of atoms located
(trapped) at different positions. In fact, the model corresponds to a
very specific geometrical configuration of the atoms confined to a
volume much smaller compared with the atomic resonant wavelength (the
small-sample model). The present atom trapping and cooling
techniques can trap two atoms at distances of order of a resonant
wavelength~\cite{eich,deb,tos}, which makes questionable the
applicability of the Dicke model to physical systems.

Recently, we have shown~\cite{ft03} that spontaneous
emission from two spatially separated atoms can lead to a transient
entanglement of initially unentangled atoms. This result contrasts the
with the Dicke model where spontaneous emission cannot produce entanglement
from initially unentangled atoms~\cite{klak,bash}.
We have numerically calculated the evolution of the
concurrence and discussed the role of the maximally entangled
collective states of the two-atom system: the rapidly decaying
symmetric state and the slowly decaying antisymmetric state of the
two-atom system. 
 
In this paper we extend our study of spontaneously induced
transient entanglement in a system of two atoms separated by an
arbitrary distance $r_{12}$. We find analytical results for the two
calculable measures of entanglement {\em concurrence} and {\em negativity}
establishing the relation between the two. Our solutions are valid for
a broad class of initial conditions including mixed states. It is
shown that when the dipole-dipole interaction becomes larger than the
atomic decay rate then the entanglement exhibits oscillatory behaviour,
oscillating with twice the frequency describing the dipole-dipole
interaction, which is the frequency separation between the symmetric and
antisymmetric states of the two-atom system. Remarkable amounts of
entanglement can be obtained at the maxima of the oscillations. For
times longer that the decay rate of the superradiant symmetric state,
when the population of the symmetric state is negligible, the
oscillations disappear and the entanglement remaining in the system is
related to the population of the slowly decaying antisymmetric state. 

\section{Measures of entanglement}
To assess how much entanglement is stored in a given quantum system it
is essential to have appropriate measures of entanglement. A number of
measures have been proposed, which include entanglement of
formation~{\cite{woo}}, entanglement of distillation~\cite{bbpssw96},
relative entropy of entanglement~\cite{vpjk97} and
negativity~\cite{per,horo,zhsl98,vw02}. For pure states, the Bell
states represent 
maximally entangled states, but for mixed states represented by a
density matrix there are some difficulties with ordering the states
according to various entanglement measures; different entanglement
measures can give different orderings of pairs of mixed states and
there is a problem of the definition of the maximally entangled mixed
state~\cite{ih00,wngkmv03}. 

Here we use two entanglement measures, {\em i.e.}, concurrence and
negativity to describe the amount of entanglement created in a
two-atom system during spontaneous emission. The concurrence
introduced by Wootters~\cite{woo} is defined as 
\begin{equation}
  \label{eq:concurrence}
  {\cal
  C}=\max\left(0,\sqrt{\lambda_{1}}-\sqrt{\lambda_{2}}-\sqrt{\lambda_{3}}-\sqrt{\lambda_{4}}\right)\, ,
\end{equation}
where $\{\lambda_{i}\}$ are the the eigenvalues of the matrix
\begin{equation}
  \label{eq:R}
  R=\rho\tilde{\rho}
\end{equation}
with $\tilde{\rho}$ given by
\begin{equation}
  \label{eq:rhot}
  \tilde{\rho}=\sigma_{x}\otimes\sigma_{x}\,\rho^{*}\,
\sigma_{x}\otimes\sigma_{x}\, , 
\end{equation}
$\sigma_{x}$ is the Pauli matrix, and $\rho$ is the density matrix
representing the quantum state.
The range of concurrence is from 0 to 1. For unentangled atoms ${\cal
C}=0$ whereas ${\cal C}=1$ for the maximally entangled atoms.

Another measure of entanglement we use here is the negativity, which
is based on the Peres-Horodecki~\cite{per,horo} criterion for
entanglement and is defined by the formula
\begin{equation}
  \label{eq:negativity}
  {\cal N}=\max\left(0,-2\sum_{i}\mu_{i}\right)\, ,
\end{equation}
where the sum is taken over the negative eigenvalues $\mu_{i}$ of the
partial transposition of the density matrix $\rho$ of the system. The
partial transposition means transposition with respect to the one atom
only. For pure states, like in the case of concurrence, ${\cal N}=1$
for maximally entangled state and ${\cal N}=0$ for unentangled atoms.

The two entanglement measures, {\em i.e.}, concurrence and negativity,
give the same criteria for entanglement, but generally they
give different values for a degree of entanglement~\cite{wngkmv03}. We
will give analytical expressions for both of them for the entanglement
produced in spontaneous emission.

Introducing the computational basis for the two-atom system as product
states of the individual atoms, as follows
\begin{eqnarray}
  \label{eq:basis}
  |1\rangle=|g_{1}\rangle\otimes|g_{2}\rangle\ ,\nonumber\\
|2\rangle=|e_{1}\rangle\otimes|e_{2}\rangle\ ,\nonumber\\
|3\rangle=|g_{1}\rangle\otimes|e_{2}\rangle\ ,\nonumber\\
|4\rangle=|e_{1}\rangle\otimes|g_{2}\rangle\ ,
\end{eqnarray}
where $|g_{i}\rangle$ and $|e_{i}\rangle$ (for $i=1,2$) are the ground
and excited states of the individual atoms, we can define the density
matrix of the two-atom system as a $4\times 4$ matrix.

We assume that the density matrix of the system has the block form
\begin{equation}
  \label{eq:rho}
  \rho=\left(
    \begin{array}[h]{cccc}
\rho_{11}&\rho_{12}&0&0\\
\rho_{21}&\rho_{22}&0&0\\
0&0&\rho_{33}&\rho_{34}\\
0&0&\rho_{43}&\rho_{44}
    \end{array}\right)
\end{equation}
with the condition $\Tr{\rho}=1$.
We will show that, if the atoms initially start from a state described
by the density matrix of the form~(\ref{eq:rho}), the evolution does not
destroy this form in the sense that the blocks of zeros remain
untouched. The other matrix elements evolve in time, and we find
explicitly their time dependence.

The matrix $\tilde{\rho}$, needed for calculation of the concurrence, has
the form
\begin{equation}
  \label{eq:rhot}
\tilde{\rho}=\left(
    \begin{array}[h]{cccc}
\rho_{22}&\rho_{12}&0&0\\
\rho_{21}&\rho_{11}&0&0\\
0&0&\rho_{44}&\rho_{34}\\
0&0&\rho_{43}&\rho_{33}
    \end{array}\right)
\end{equation}
and the square roots of the eigenvalues of the matrix $R$ given
by~(\ref{eq:R}) are the following
\begin{eqnarray}
  \label{eq:roots}
\fl \left\{\sqrt{\lambda_{i}}\right\}=&&\left\{
\sqrt{\rho_{11}\rho_{22}}-|\rho_{12}|,
\sqrt{\rho_{11}\rho_{22}}+|\rho_{12}|,\sqrt{\rho_{33}\rho_{44}}-|\rho_{34}|,
\sqrt{\rho_{33}\rho_{44}}+|\rho_{34}|\,\right\}\, .
\end{eqnarray}
Depending on the particular values of the matrix elements there are
two possibilities for the largest eigenvalue, either the second term or
the fourth term in~(\ref{eq:roots}). 
The concurrence is thus given by
\begin{equation}
  \label{eq:concurrence1}
  {\cal C}=
\max\left\{0,\,{\cal C}_{1},\,{\cal C}_{2}
\right\},
\end{equation}
with
\begin{eqnarray}
  \label{eq:altconc}
  {\cal C}_{1}=2\,(|\rho_{12}|-\sqrt{\rho_{33}\rho_{44}}\,)\ ,\nonumber\\
{\cal C}_{2}=2\,(|\rho_{34}|-\sqrt{\rho_{11}\rho_{22}}\,)\ ,
\end{eqnarray}
and we have two alternative expressions for the concurrence depending
on which of them is positive.

For calculation of the negativity we need the partially transposed
density matrix. The transposition with respect to the indices of the
first atom gives the matrix
\begin{equation}
  \label{eq:transposed}
  \rho^{{\rm T}_{1}}=\left(
    \begin{array}[h]{cccc}
\rho_{11}&\rho_{43}&0&0\\
\rho_{34}&\rho_{22}&0&0\\
0&0&\rho_{33}&\rho_{21}\\
0&0&\rho_{12}&\rho_{44}
    \end{array}\right)
\end{equation}
which has the eigenvalues
\begin{eqnarray}
  \label{eq:eigtransp}
  \left\{\nu_{i}\right\}=&&
\left\{\frac{1}{2}\left(\rho_{11}+\rho_{22}\pm\sqrt{(\rho_{11}+\rho_{22})^{2}
+4\,(|\rho_{34}|^{2}-\rho_{11}\rho_{22})}\right),\right.\nonumber\\
&&\left.\frac{1}{2}\left(\rho_{33}+\rho_{44}\pm\sqrt{(\rho_{33}+\rho_{44})^{2}
+4\,(|\rho_{12}|^{2}-\rho_{33}\rho_{44})}\right)\right\}\ .
\end{eqnarray}
There are two candidates for being negative among the
roots~(\ref{eq:eigtransp}), however, they cannot be negative
simultaneously because the inequality
$|\rho_{34}|-\sqrt{\rho_{11}\rho_{22}}>0$ implies that
$|\rho_{12}|-\sqrt{\rho_{33}\rho_{44}}<0$, and vice versa. 
It is also easy to find the two alternative values for the
concurrence~(\ref{eq:concurrence1}) inside the square roots.
So, the negativity defined by~(\ref{eq:negativity}) has also two
alternative forms 
\begin{eqnarray}
  \label{eq:negativity1}
  {\cal N}=
\max\left\{0,\,\sqrt{4\,(|\rho_{12}|^{2}
-\rho_{33}\rho_{44})+(\rho_{33}+\rho_{44})^{2}}-(\rho_{33}+\rho_{44})
\right.,\nonumber\\
\left.\phantom{\max\qquad\qquad}\sqrt{4\,(|\rho_{34}|^{2}
-\rho_{11}\rho_{22})+(\rho_{11}+\rho_{22})^{2}}-(\rho_{11}+\rho_{22})
\right\},\\
\label{eq:negativity1_2}
\phantom{{\cal N}}=
\max\left\{0,\sqrt{{\cal C}_{1}\,{\cal C}^{+}_{1}\,+(\rho_{33}+\rho_{44})^{2}}
-(\rho_{33}+\rho_{44})\right.,\nonumber\\
\left.\phantom{\max\qquad\qquad}\sqrt{{\cal C}_{2}\,{\cal C}^{+}_{2}\,
+(\rho_{11}+\rho_{22})^{2}}-(\rho_{11}+\rho_{22})\right\},
\end{eqnarray}
where the appropriate expression from~(\ref{eq:altconc}) is to be
substituted to the two alternative terms in~(\ref{eq:negativity1_2}).
The quantities ${\cal C}^{+}_{1}$, and ${\cal C}^{+}_{2}$, which are
always nonnegative, represent two alternative expressions associated
with corresponding expressions ${\cal C}_{1}$ and ${\cal C}_{2}$ for
the concurrence~(\ref{eq:concurrence1}), and they have the following form
\begin{eqnarray}
  \label{eq:Cplus}
  {\cal C}^{+}_{1}=
2\,(|\rho_{12}|+\sqrt{\rho_{33}\rho_{44}}\,),\nonumber\\
{\cal C}^{+}_{2}=2\,(|\rho_{34}|+\sqrt{\rho_{11}\rho_{22}}\,).
\end{eqnarray}
For pure states ${\cal C}_{1}^{+}$ and ${\cal C}_{2}^{+}$ are equal to 
 ${\cal C}_{1}$ and ${\cal C}_{2}$, respectively, and in this case the
 negativity is equal to the concurrence.

The equality~(\ref{eq:negativity1_2}) establishes the relation
between the negativity and the concurrence for the system described by
the density matrix of the form~(\ref{eq:rho}). It is evident that both
quantities give the same criterion for entanglement, that is, positive
value of ${\cal C}$ implies positive value of ${\cal N}$, but the
degree of entanglement indicated by the two quantities can be quite
different. It is also clear from~(\ref{eq:negativity1_2}) that the
product ${\cal C}_{1}\,{\cal C}_{1}^{+}$ (${\cal C}_{2}\,{\cal
  C}_{2}^{+}$) can as a whole serve as a
measure of entanglement: it is zero if ${\cal C}=0$, and it is unity
for maximally entangled pure state, but for mixed states it gives yet
another value for the degree of entanglement.

It is interesting to express the results for concurrence and
negativity in the Bell basis which is defined as follows
\begin{eqnarray}
|1'\rangle=|\Phi^{+}\rangle=\phantom{-}\frac{1}{\sqrt{2}}\left(|1\rangle
+|2\rangle\right)\ ,\nonumber\\
|2'\rangle=|\Phi^{-}\rangle=-\frac{1}{\sqrt{2}}\left(|1\rangle
-|2\rangle\right)\ ,\nonumber\\
|3'\rangle=|\Psi^{+}\rangle=\phantom{-}\frac{1}{\sqrt{2}}\left(|3\rangle
+|4\rangle\right)\ ,\nonumber\\
|4'\rangle=|\Psi^{-}\rangle=-\frac{1}{\sqrt{2}}\left(|3\rangle
-|4\rangle\right)\, .
\label{bellbasis}
\end{eqnarray}
The transformation of the density matrix $\rho$ given
by~(\ref{eq:rho}) from the original basis~(\ref{eq:basis}) to the Bell
basis~(\ref{bellbasis}) is performed with the transformation matrix
\begin{eqnarray}
\label{tobellU}
U=\frac{1}{\sqrt{2}}\left(
\begin{array}[h]{rrrr}
1&1&0&0\\
-1&1&0&0\\
0&0&1&1\\
0&0&-1&1\\
\end{array}\right)
\end{eqnarray}
leading to the new density matrix
\begin{eqnarray}
\label{tobell}
\rho'=U\rho U^{+}\ ,
\end{eqnarray}
which has the same block form as~(\ref{eq:rho}) but the new matrix
elements have primed indices ($\rho_{1'1'},\rho_{1'2'},\dots$).
The matrix elements in the Bell basis are related to the original
matrix elements as follows
\begin{eqnarray}
  \label{eq:bellelements}
\fl  \begin{array}[h]{ll}
\rho_{1'1'}=\phantom{-}\frac{1}{2}\left[\rho_{11}+\rho_{22}
+(\rho_{12}+\rho_{21})\right],
&\qquad\rho_{3'3'}=\phantom{-}\frac{1}{2}\left[\rho_{33}+\rho_{44}
+(\rho_{12}+\rho_{21})\right]\ ,\\
\rho_{2'2'}=\phantom{-}\frac{1}{2}\left[\rho_{11}+\rho_{22}
-(\rho_{12}+\rho_{21})\right],
&\qquad\rho_{4'4'}=\phantom{-}\frac{1}{2}\left[\rho_{33}+\rho_{44}
-(\rho_{34}+\rho_{43})\right]\ ,\\
\rho_{1'2'}=-\frac{1}{2}\left[\rho_{11}-\rho_{22}
+(\rho_{12}-\rho_{21})\right],
&\qquad\rho_{3'4'}=-\frac{1}{2}\left[\rho_{33}-\rho_{44}
+(\rho_{34}-\rho_{43})\right]\ ,\\
\rho_{2'1'}=-\frac{1}{2}\left[\rho_{11}-\rho_{22}
-(\rho_{12}-\rho_{21})\right],
&\qquad\rho_{4'3'}=-\frac{1}{2}\left[\rho_{33}-\rho_{44}
-(\rho_{34}-\rho_{43})\right]\ .
  \end{array}
\end{eqnarray}
In the Bell basis~(\ref{bellbasis}),
 the concurrence alternatives~(\ref{eq:altconc}) and  
the negativity~(\ref{eq:negativity1_2}) take the following form
\begin{eqnarray}
  \label{eq:concurrence2}
\fl  {\cal C}_{1}=
\sqrt{(\rho_{1'1'}-\rho_{2'2'})^{2}-(\rho_{1'2'}
-\rho_{2'1'})^{2}}
 -\sqrt{(\rho_{3'3'}+\rho_{4'4'})^{2}-(\rho_{3'4'}
+\rho_{4'3'})^{2}}\ ,\nonumber\\
\fl{\cal C}_{2}=\sqrt{(\rho_{3'3'}-\rho_{4'4'})^{2}-(\rho_{3'4'}
-\rho_{4'3'})^{2}}
-\sqrt{(\rho_{1'1'}+\rho_{2'2'})^{2}-(\rho_{1'2'}+\rho_{2'1'})^{2}}
\end{eqnarray}
\begin{eqnarray}
  \label{eq:negativity2}
  {\cal N}=
\max\left\{0,\sqrt{{\cal C}_{1}\,{\cal C}^{+}_{1}\,+(\rho_{3'3'}
+\rho_{4'4'})^{2}}-(\rho_{3'3'}+\rho_{4'4'})\right.,\nonumber\\
\left.\phantom{\max\qquad\qquad}\sqrt{{\cal C}_{2}\,{\cal C}_{2}^{+}
+(\rho_{1'1'}+\rho_{2'2'})^{2}}-(\rho_{1'1'}+\rho_{2'2'})
\right\}\ ,
\end{eqnarray}
where
\begin{eqnarray}
  \label{eq:Cplus1}
\fl  {\cal C}^{+}_{1}=
\sqrt{(\rho_{1'1'}-\rho_{2'2'})^{2}-(\rho_{1'2'}-\rho_{2'1'})^{2}}
 +\sqrt{(\rho_{3'3'}+\rho_{4'4'})^{2}-(\rho_{3'4'}
+\rho_{4'3'})^{2}}\ ,\nonumber\\
\fl{\cal C}^{+}_{2}=\sqrt{(\rho_{3'3'}-\rho_{4'4'})^{2}-(\rho_{3'4'}
-\rho_{4'3'})^{2}}+\sqrt{(\rho_{1'1'}+\rho_{2'2'})^{2}-(\rho_{1'2'}
+\rho_{2'1'})^{2}}\ .
\end{eqnarray}
From~(\ref{eq:concurrence2}) and~(\ref{eq:negativity2}), it is evident
that for any Bell state~(\ref{bellbasis}) the concurrence and negativity
become unity. For mixed states the situation is much more complicated
with the values of concurrence and negativity which are different in
this case, and have values between zero and unity. Later on we 
apply the general 
formulas derived in this Section to find the evolution of entanglement
in spontaneous emission from a system of two two-level atoms. 

\section{Atomic evolution}
We consider a system of two non-overlapping two-level atoms with
ground states $\left|g_{i}\right\rangle$ and excited states
$\left|e_{i}\right\rangle \ (i=1,2)$ connected by dipole transition
moments $\vec{\mu}_{i}$. The atoms are
located at fixed positions $\vec{r}_{1}$ and $\vec{r}_{2}$
and coupled to all modes of the electromagnetic field, which we assume
are in the vacuum state. We consider spontaneous emission from identical
as well as non-identical atoms prepared in  different initial states.
In the case of nonidentical atoms, we assume that atoms have equal dipole
moments
$\mbox{\boldmath$\mu$}_{1}=\mbox{\boldmath$\mu$}_{2}=\mbox{\boldmath$\mu$}$,
but 
different transition frequencies $\omega_{1}$ and
$\omega_{2}$, such that $\omega_{2}-\omega_{1} \ll \omega_{0}=
(\omega_{1}+\omega_{2})/2$, so that the rotating-wave approximation
can be applied to calculate the dynamics of the system.

The time evolution of the system of atoms coupled through the vacuum
field is given by the following master equation~\cite{leh,ag74,ftk87}
\begin{eqnarray}
      \frac{\partial {\rho}}{\partial t} &=&
      -i\sum_{i=1}^{2}\omega_{i}
      \left[S^{z}_{i},{\rho}\right]
      -i\sum_{i\neq j}^{2}\Omega_{ij}
      \left[S^{+}_{i}S^{-}_{j},{\rho}\right] \nonumber \\
      &-& \frac{1}{2}\sum_{i,j=1}^{2}\Gamma _{ij}\left( {\rho}
      S_{i}^{+}S_{j}^{-}+S_{i}^{+}S_{j}^{-}{\rho}
      -2S_{j}^{-}{\rho} S_{i}^{+}\right) \ , 
\label{eq:mastereq}
\end{eqnarray}
where $S_{i}^{+}\ (S_{i}^{-})$ are the dipole raising (lowering)
operators and $S^{z}$ is the energy operator of the $i$th atom. In
Eq.~(\ref{eq:mastereq}), $\Gamma_{ij}\ (i=j)$ are the spontaneous
emission rates of the atoms, equal to the Einstein $A$ coefficient
for spontaneous emission, whereas $\Gamma_{ij}$ and $\Omega_{ij}\
(i\neq j)$ describe the interatomic coupling~\cite{leh,ag74,ftk87},
and are the collective damping and the dipole-dipole interaction
potential defined, respectively, by
\begin{eqnarray}
\Gamma_{ij}=\Gamma_{ji}&=&
\frac{3}{2}\Gamma
\left\{ \left[1 -\left( \hat{\mbox{\boldmath$\mu$}}\cdot
      \hat{\mbox{\boldmath$r$}} 
_{ij}\right)^{2} \right] \frac{\sin \left( k_{0}r_{ij}\right)
}{k_{0}r_{ij}}\right.  \nonumber \\
&&\left. +\left[ 1 -3\left( \hat{\mbox{\boldmath$\mu$}}\cdot
\hat{\mbox{\boldmath$r$}}_{ij}\right)^{2} \right] \left[ \frac{\cos \left(
k_{0}r_{ij}\right) }{\left( k_{0}r_{ij}\right) ^{2}}-\frac{\sin \left(
k_{0}r_{ij}\right) }{\left( k_{0}r_{ij}\right) ^{3}}\right] \right\} \ ,
\label{eq:Gamma}
\end{eqnarray}
and
\begin{eqnarray}
\Omega_{ij}&=&
\frac{3}{4}\Gamma
\left\{ -\left[1 -\left( \hat{\mbox{\boldmath$\mu$}}\cdot
      \hat{\mbox{\boldmath$r$}} 
_{ij}\right)^{2} \right] \frac{\cos \left( k_{0}r_{ij}\right)
}{k_{0}r_{ij}}\right.  \nonumber \\
&&\left. +\left[ 1 -3\left( \hat{\mbox{\boldmath$\mu$}}\cdot
\hat{\mbox{\boldmath$r$}}_{ij}\right)^{2} \right] \left[ \frac{\sin \left(
k_{0}r_{ij}\right) }{\left( k_{0}r_{ij}\right) ^{2}}+\frac{\cos \left(
k_{0}r_{ij}\right) }{\left( k_{0}r_{ij}\right) ^{3}}\right] \right\} \ ,
\label{eq:Omega}
\end{eqnarray}
where $k_{0}=\omega_{0}/c$, $r_{ij}
=\left|\mbox{\boldmath$r$}_{j}-\mbox{\boldmath$r$}_{i}\right|$ 
is the distance between the atoms, $\bar{\mu}$ is unit vector
along the atomic transition dipole moments, that we assume are
parallel to each other, and $\bar{r}_{ij}$ is the unit vector along
the interatomic axis.

The master equation~(\ref{eq:mastereq}) has been used for many years to study
a wide variety of problems involving the interaction of collective
atomic systems with the radiation field~\cite{ft02}.
Using the master equation~(\ref{eq:mastereq}), we can write down the equations
of motion for the components of the density matrix of the two-atom
system in the basis~(\ref{eq:basis}) of the product states
\begin{figure}[htb]
  \centering
  \includegraphics[height=6cm]{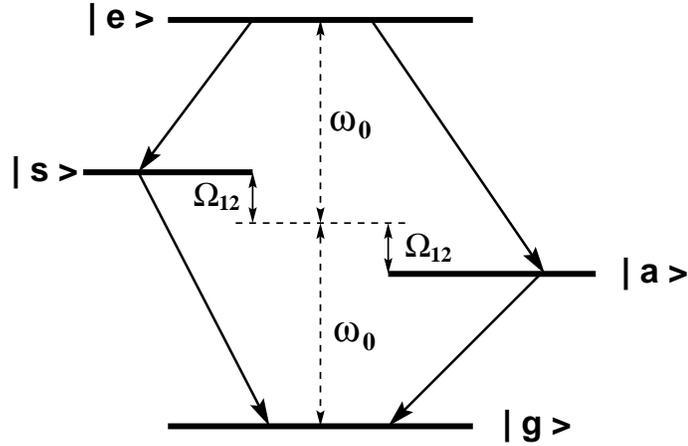}
  \caption{Collective states of two identical atoms}
  \label{fig:colstat}
\end{figure}
of the individual atoms. However, the problem simplifies by working in
the basis of the 
collective states of the system which contains symmetric and
antisymmetric combinations of the product states. For identical atoms
$(\omega_{1}=\omega_{2})$ the collective states are~\cite{dic,leh}
\begin{eqnarray}
      |g\rangle &=& |1\rangle \ ,\nonumber\\
      |e\rangle &=& |2\rangle\ ,\nonumber \\
      |s\rangle &=& \frac{1}{\sqrt{2}}
      \left(|3\rangle +|4\rangle\right) \ ,\nonumber \\
      |a\rangle &=& \frac{1}{\sqrt{2}}
      \left(|4\rangle-|3\rangle\right) \ , 
\label{eq:atomstates}
\end{eqnarray}
where we used the basis~(\ref{eq:basis}).

In the collective state representation, the two-atom system behaves
as a single four-level system, illustrated in Fig.~\ref{fig:colstat}, with
the ground state 
$\left|g\right\rangle$, the upper state $\left|e\right\rangle$, and
two intermediate states: the symmetric $\left|s\right\rangle$ and
antisymmetric $\left|a\right\rangle$ states. The most important
property of the collective states is that the symmetric and
antisymmetric states are maximally entangled states. The states are
linear superpositions of the product states which cannot be separated
into product states of the individual atoms. They are in fact two of the
Bell states introduced in~(\ref{bellbasis}): $|s\rangle=|3'\rangle$
and $|a\rangle=|4'\rangle$. The symmetric and antisymmetric states are
eigenstates of the system of two identical atoms with the
dipole-dipole interaction included. The basis of atomic
states~(\ref{eq:atomstates}) can be considered as an effect of partial
transformation to the Bell basis~(\ref{bellbasis}) in which the
transformation has been performed in the lower block only. Such basis
is convenient for finding the solution to the master
equation~(\ref{eq:mastereq}) describing spontaneous emission in the
system. 

Assuming that initially the state of the system has been prepared in
the block form~(\ref{eq:rho}), from the master
equation~(\ref{eq:mastereq}), we get the following set of differential
equations 
describing the evolution of the system in the basis of collective atomic
states~(\ref{eq:atomstates}) 
\begin{eqnarray}
      \dot{\rho}_{ee} &=& -2\Gamma \rho_{ee} \ ,\nonumber \\
       \dot{\rho}_{eg} &=& -\left(\Gamma
       +2i\omega_{0}\right)\rho_{eg}\ ,\nonumber\\
      \dot{\rho}_{ss} &=& -\left(\Gamma
      +\Gamma_{12}\right)\left(\rho_{ss} -\rho_{ee}\right)
      +i\Delta \left(\rho_{as}-\rho_{sa}\right) \ ,\nonumber \\
      \dot{\rho}_{aa} &=& -\left(\Gamma
      -\Gamma_{12}\right)\left(\rho_{aa} -\rho_{ee}\right)
      -i\Delta \left(\rho_{as}-\rho_{sa}\right) \ ,\nonumber \\
      \dot{\rho}_{as} &=& -\left(\Gamma +2i\Omega_{12}\right)\rho_{as}
      +i\Delta \left(\rho_{ss}-\rho_{aa}\right) \ 
      \label{eq:eqsofmotion}
\end{eqnarray}
with the condition $\rho_{gg}+\rho_{ee}+\rho_{ss}+\rho_{aa}=1$, and
with $\Delta=(\omega_{2}-\omega_{1})/2$. All
other matrix elements related to the blocks of zeros in~(\ref{eq:rho})
remain zeros, if the evolution is govern by the master
equation~(\ref{eq:mastereq}). 

Equations~(\ref{eq:eqsofmotion}) show that all transitions rates to
and from the 
symmetric state are equal to $(\Gamma +\Gamma_{12})$. On the other
hand, all transitions rates to and from the antisymmetric state are
equal to $(\Gamma -\Gamma_{12})$. Thus, the symmetric state decays
with an enhanced (superradiant) rate, while the antisymmetric state
decays with a reduced (subradiant) state. Hence, the population of
the antisymmetric state experiences a variation on a time scale of
order $(\Gamma -\Gamma_{12})^{-1}$, which can lead to interesting
effects not observed in the Dicke model. These effects result from
the fact that the set of equations~(\ref{eq:eqsofmotion}) has two
different solutions 
depending on whether $\Gamma_{12}=\Gamma$ or $\Gamma_{12}\neq
\Gamma$. The case of $\Gamma_{12}=\Gamma$ corresponds to the small
sample (Dicke) model, whereas the case of $\Gamma_{12}\neq \Gamma$
corresponds to spatially extended atomic systems. The existence of
two different solutions of Eq.~(\ref{eq:eqsofmotion}) is connected
with conservation 
of the total spin $S^{2}$, that $S^{2}$ is a constant of motion for the
Dicke model and $S^{2}$ not being a constant of motion for a spatially
extended system of atoms~\cite{ftk81,hsf82}. We can explain it by
expressing the square of the total spin of the two-atom system in
terms of the density matrix elements of the collective system as
\begin{eqnarray}
         S^{2}\left(t\right) = 2 -2\rho_{aa}\left(t\right) \ .
\label{eq:S2}
\end{eqnarray}
It is clear from Eq.~(\ref{eq:S2}) that $S^{2}$ is conserved only in
the Dicke model, in which the antisymmetric state is ignored. For a
spatially extended system the antisymmetric state participates fully
in the dynamics and $S^{2}$ is not conserved. The Dicke model evolves
between the triplet states $\left|e\right\rangle$, $\left|s\right\rangle$,
and $\left|g\right\rangle$, while the spatially extended two-atom system
evolves between the triplet and the antisymmetric states.

The problem of spontaneous emission from two atoms can be solved
analytycally even for general case of nonidentical atoms ($\Delta\ne
0$)\cite{ftk86}, but 
the general solutions are rather lengthy, and we will give here the
solutions for the simpler case of identical atoms only. It is seen
from~(\ref{eq:eqsofmotion}) that the first two equations, belonging to
the upper block of~(\ref{eq:rho}), are decoupled from the other
equations belonging to the lower block of~(\ref{eq:rho}), and they
have simple exponential solutions
\begin{eqnarray}
  \label{eq:sol1}
  \rho_{ee}(t)=\rho_{ee}(0)\,\e^{-2\Gamma t}\ ,\nonumber\\
\rho_{eg}(t)=\rho_{eg}(0)\,\e^{-(\Gamma+2 i\omega_{0})t}\ .
\end{eqnarray}
For identical atoms, $\Delta=0$, the remaining equations simplify
considerably and their solutions are as follows
\begin{eqnarray}
  \label{eq:sol2}
  \rho_{ss}(t)=\rho_{ss}(0)\,\e^{-(\Gamma+\Gamma_{12})t}+\rho_{ee}(0)\,
\frac{\Gamma+\Gamma_{12}}{\Gamma-\Gamma_{12}}
\left(\e^{-(\Gamma+\Gamma_{12})t}-\e^{-2\Gamma t}\right)\ ,\nonumber\\
\rho_{aa}(t)=\rho_{aa}(0)\,\e^{-(\Gamma-\Gamma_{12})t}
+\rho_{ee}(0)\,\frac{\Gamma-\Gamma_{12}}{\Gamma+\Gamma_{12}}
\left(\e^{-(\Gamma-\Gamma_{12})t}-\e^{-2\Gamma t}\right)\ ,\nonumber\\
\rho_{as}(t)=\rho_{as}(0)\,\e^{-(\Gamma+2i\Omega_{12})t}\ .
\end{eqnarray}
The evolution within the two blocks runs independently except for the
fact that all the states decay to the ground state $|g\rangle$, and
the population of this state is
\begin{eqnarray}
  \label{eq:ground}
  \rho_{gg}(t)=1-\rho_{ee}(t)-\rho_{ss}(t)-\rho_{aa}(t)\ ,
\end{eqnarray}
{\em i.e}, eventually total atomic population accumulates in the
ground state.

The solutions~(\ref{eq:sol1}) and~(\ref{eq:sol2}) are particularly
simple in the basis~(\ref{eq:atomstates}), but it is quite easy to
transform them into the original basis~(\ref{eq:basis}) or into the
Bell basis~(\ref{bellbasis}) using the
relations~(\ref{eq:bellelements}). The same transformation can be used
to transform the initial conditions. In this way we obtain analytical
results for the matrix elements of the density matrix in either the
original basis or the Bell basis as the linear combinations of the
solutions~(\ref{eq:sol1}) and~(\ref{eq:sol2}), for any initial
conditions that preserve the block form~(\ref{eq:rho}) of the density matrix.

\section{Entanglement in the two-atom system}
The solutions obtained in the previous Section can be used in
formulas~(\ref{eq:concurrence1}), (\ref{eq:altconc})
or~(\ref{eq:concurrence2}) for the concurrence and in
formulas~(\ref{eq:negativity1}) or~(\ref{eq:negativity2}) for the
negativity giving the analytical expressions for the quantities
describing degree of entanglement in the system. For example, if
$\rho_{eg}(0)=\rho_{21}(0)=0$, {\em i.e.}, there is no two-photon
coherence in the system initially, then ${\cal C}_{1}$ cannot be
positive, so it cannot contribute to the concurrence ${\cal C}$, and the
concurrence is equal to ${\cal C}_{2}$, if it is positive. We have
\begin{eqnarray}
  \label{eq:concurrence3}
  {\cal C}(t)=\max\left\{0,{\cal C}_{2}(t)\right\}\ ,
\end{eqnarray}
and
\begin{eqnarray}
  \label{eq:concurrence4}
  {\cal~C}_{2}(t)=\sqrt{\left[\rho_{ss}(t)-\rho_{aa}(t)\right]^{2}
-\left[\rho_{sa}(t)-\rho_{as}(t)\right]^{2}}-2\sqrt{\rho_{ee}(t)\rho_{gg}(t)}
\end{eqnarray}
with the solutions~(\ref{eq:sol1})--(\ref{eq:ground}). The
solution~(\ref{eq:concurrence4}) still covers a broad range of
initial conditions, {\em i.e.}, such that the upper block
in~(\ref{eq:rho}) is diagonal 
but the lower block is arbitrary. 

It is immediately seen from~(\ref{eq:concurrence4}) and the
solutions~(\ref{eq:sol2}) that, for two identical atoms prepared
initially in one of the maximally entangled states $|s\rangle$ or
$|a\rangle$, the concurrence for any time is equal to the population of
the corresponding state $\rho_{ss}(t)$ or $\rho_{aa}(t)$: it is unity
at time $t=0$ 
and decays in time at rate $\Gamma+\Gamma_{12}$ for the symmetric
state and at rate $\Gamma-\Gamma_{12}$ for the antisymmetric state.

The quantity
${\cal C}_{2}^{+}$, defined by~(\ref{eq:Cplus}), associated with
${\cal~C}_{2}$ is then given by 
\begin{eqnarray}
  \label{eq:C2plus}
  {\cal C}_{2}^{+}(t)=\sqrt{\left[\rho_{ss}(t)-\rho_{aa}(t)\right]^{2}
-\left[\rho_{sa}(t)-\rho_{as}(t)\right]^{2}}+2\sqrt{\rho_{ee}(t)
\rho_{gg}(t)}\ , 
\end{eqnarray}
and the negativity ${\cal N}$, given by~(\ref{eq:negativity1_2}),
evolves in time according to the formula
\begin{eqnarray}
  \label{eq:negatC2plus}
  {\cal N}(t)=\max\left\{0,{\cal N}_{2}(t)\right\}\ ,
\end{eqnarray}
where
\begin{eqnarray}
  \label{eq:N2}
{\cal N}_{2}(t)=\sqrt{{\cal C}_{2}(t)\,{\cal C}_{2}^{+}(t)
+\left[\rho_{gg}(t)+\rho_{ee}(t)\right]^{2}}
-\left[\rho_{gg}(t)+\rho_{ee}(t)\right]\ .
\end{eqnarray}
Equations~(\ref{eq:concurrence4}) and~(\ref{eq:N2}) are exact
analytical formulas describing the time evolution of entanglement
created in the system of two identical atoms via the process of spontaneous
emission. 

Let us now consider two special cases of the initial conditions:
(i) initially only one atom excited, $\rho_{44}(0)=1$, (ii) both atoms
initially excited, $\rho_{ee}=1$. 

For case (i), we have
$\rho_{ss}(0)=\rho_{aa}(0)=\rho_{as}(0)=\rho_{sa}(0)=1/2$,
$\rho_{ee}(0)=0$, and equation~(\ref{eq:concurrence4}) takes the form
\begin{eqnarray}
  \label{eq:C2one}
  {\cal
  C}_{2}(t)=\frac{1}{2}\sqrt{\left[\e^{-(\Gamma+\Gamma_{12})t}
-\e^{-(\Gamma-\Gamma_{12})t}\right]^{2}
+\e^{-2\Gamma t}\sin^{2}(2\Omega_{12}t)}\ .
\end{eqnarray}
From~(\ref{eq:C2one}) it is seen that ${\cal C}_{2}(0)=0$, there is no
entanglement at $t=0$, as it should be since the initial state is a
product state. However, for $t>0$, ${\cal C}_{2}(t)$ becomes positive,
which means that the two atoms become entangled, and the degree of
entanglement measured by the concurrence is given
by~(\ref{eq:C2one}). For long times, all terms in~(\ref{eq:C2one}) decay
to zero, and the concurrence goes to zero. One more interesting
feature of the evolution is seen from~(\ref{eq:C2one}), it is the
oscillatory behaviour of the concurrence which can be observed at
times shorter than $(2\Gamma)^{-1}$, when the oscillatory term
contributes significantly to the evolution. The oscillations are with
the frequency $2\Omega_{12}$, which is equal to the separation of the
symmetric and antisymmetric states, and the oscillations become visible
when the dipole-dipole interaction is sufficiently strong,
{\em i.e.} for $\Omega_{12}\gg \Gamma$. For times longer than
$(2\Gamma)^{-1}$ the only term that survives is the term that decays
with the rate $\Gamma-\Gamma_{12}$, which comes from the evolution of
\begin{figure}[htb]
  \centering
  \includegraphics[height=9cm]{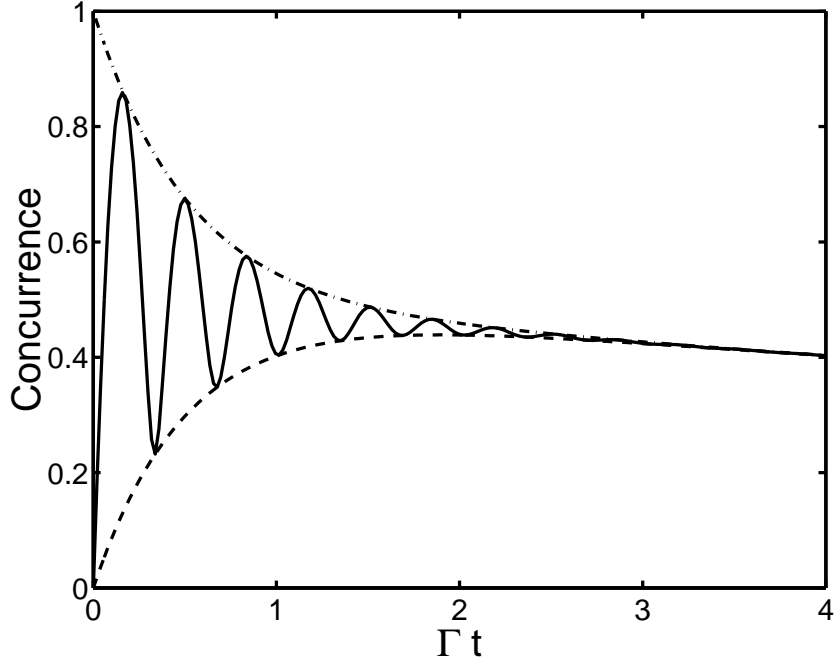}
  \caption{Time evolution of the concurrence ${\cal C}(t)$ (solid line),
  $\rho_{aa}(t)-\rho_{ss}(t)$ (dashed line), and
  $\rho_{aa}(t)+\rho_{ss}(t)$ (dashed-dotted line) for initially
one atom excited ($\rho_{44}(0)=1$) with $\mbox{\boldmath$\hat{\mu}$}
\perp\mbox{\boldmath$\hat{r}$}_{12}$, and $r_{12}=\lambda/12$
  ($\Gamma_{12}=0.95\,\Gamma$, $\Omega_{12}=9.30\,\Gamma$).}
  \label{fig:conc1}
\end{figure}
the slowly decaying antisymmetric state, and the concurrence becomes
equal to the population of this state. We have numerically
studied~\cite{ft03} this behaviour, but for the interatomic distances
not so short as to reveal the oscillations in the concurrence. 

In Fig.~\ref{fig:conc1} we present the oscillatory behaviour of the
concurrence~(\ref{eq:C2one}) for the case of initially one atom
\begin{figure}[htb]
  \centering
  \includegraphics[height=9cm]{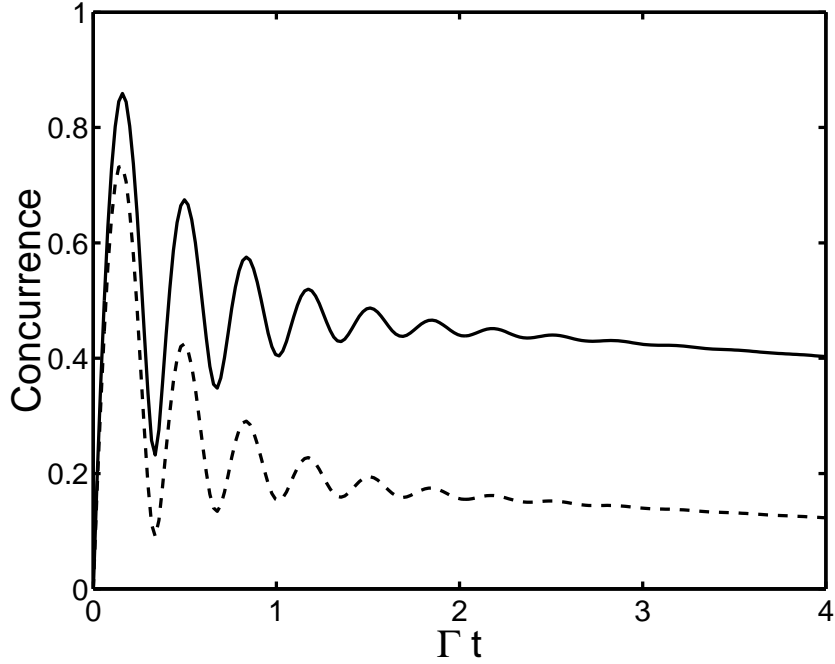}
  \caption{Comparison of concurrence (solid) and negativity (dashed)
  for the same parameters as in Fig.~\ref{fig:conc1}.}
  \label{fig:neg1}
\end{figure}
excited and the interatomic distance $r_{12}=\lambda/12$, which gives
the values $\Gamma_{12}=0.95\, \Gamma$ and
$\Omega_{12}=9.30\,\Gamma$. The envelops of the oscillations are given
by $\rho_{aa}(t)-\rho_{ss}(t)$ for the lower envelope and
$\rho_{aa}(t)+\rho_{ss}(t)$ for the upper envelope. The value of concurrence
at the first maximum is $0.86$, which is quite remarkable. After the
time $t\sim (2\Gamma)^{-1}$ when the symmetric state is practically
depopulated, the concurrence becomes equal to the population of the
antisymmetric state. 

The negativity for this case, from~(\ref{eq:C2plus})--(\ref{eq:N2}),
takes the form
\begin{eqnarray}
  \label{eq:N2one}
\fl  {\cal N}_{2}(t)=\sqrt{\frac{1}{4}\left\{\left[\e^{-(\Gamma+\Gamma_{12})t}
-\e^{-(\Gamma-\Gamma_{12})t}\right]^{2}+\e^{-2\Gamma t}\sin^{2}(2\Omega_{12}t)
\right\}+\rho_{gg}^{2}(t)}-\rho_{gg}(t)\ ,
\end{eqnarray}
where
\begin{eqnarray}
  \label{eq:rhogg}
  \rho_{gg}(t)=1-\frac{1}{2}\left[\e^{-(\Gamma+\Gamma_{12})t}
+\e^{-(\Gamma-\Gamma_{12})t}\right]\ .
\end{eqnarray}
In Fig.~\ref{fig:neg1} we compare the time evolution of the two
measures of entanglement: concurrence and negativity for the same
values of the parameters as in Fig.~\ref{fig:conc1}. Generally, the
negativity takes smaller values than the concurrence, except for the
initial value which is zero for both of them, and the value for
$t\rightarrow\infty$ which is also zero. 

For case (ii), we have $\rho_{ee}(0)=1$ and the
concurrence~(\ref{eq:concurrence4}) takes the form
\begin{eqnarray}
  \label{eq:C2two}
\fl  {\cal C}_{2}(t)=\left|\frac{\Gamma+\Gamma_{12}}{\Gamma-\Gamma_{12}}
\left(\e^{-(\Gamma+\Gamma_{12})t}-\e^{-2\Gamma t}\right)
-\frac{\Gamma-\Gamma_{12}}{\Gamma+\Gamma_{12}}
\left(\e^{-(\Gamma-\Gamma_{12})t}-\e^{-2\Gamma t}\right)\right|
-2\e^{-\Gamma t}\sqrt{\rho_{gg}}
\end{eqnarray}
with
\begin{eqnarray}
  \label{eq:rhogg1}
\fl  \rho_{gg}(t)=1-\left[\frac{\Gamma+\Gamma_{12}}{\Gamma-\Gamma_{12}}
\left(\e^{-(\Gamma+\Gamma_{12})t}-\e^{-2\Gamma t}\right)
+\frac{\Gamma-\Gamma_{12}}{\Gamma+\Gamma_{12}}
\left(\e^{-(\Gamma-\Gamma_{12})t}-\e^{-2\Gamma t}\right)
+\e^{-2\Gamma t}\right]\ .
\end{eqnarray}
Again, for $t=0$ the concurrence is zero, but now it is not easy to
see if ${\cal C}_{2}(t)$ can be positive, and numerical evaluation is
needed to check the positivity. What is clear from~(\ref{eq:C2two}),
however, it is the fact that there are no oscillations in this case.
\begin{figure}[htb]
  \centering
  \includegraphics[height=9cm]{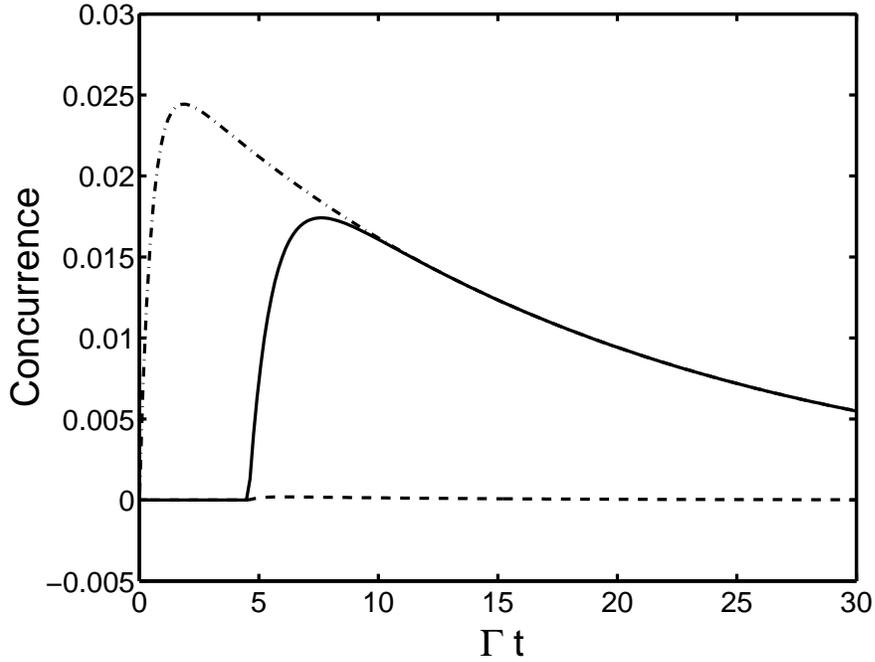}
  \caption{Time evolution of concurrence ${\cal C}(t)$ (solid),
  negativity ${\cal N}(t)$ (dashed),
  and population of the antisymmetric state $\rho_{aa}(t)$
  (dashed-dotted), for initially 
both  atoms excited ($\rho_{ee}(0)=1$) with $\mbox{\boldmath$\hat{\mu}$}
\perp\mbox{\boldmath$\hat{r}$}_{12}$, and $r_{12}=\lambda/12$
  ($\Gamma_{12}=0.95\,\Gamma$, $\Omega_{12}=9.30\,\Gamma$).}
  \label{fig:conc2}
\end{figure}
One can also expect that for times for which the populations of the
excited state and the symmetric state, which decay much faster than the
antisymmetric state, are already close to zero, it 
is still some population in the antisymmetric state and ${\cal C}_{2}(t)$
becomes positive. Numerical evaluation of~(\ref{eq:C2two}) confirm
that it is really true. Corresponding formula for the negativity can be
obtained from~(\ref{eq:C2plus}), (\ref{eq:N2}) and~(\ref{eq:C2two}),
but this are just simple substitutions, so we do not write it explicitly.
In Fig.~\ref{fig:conc2} we plot the time evolution of the concurrence
${\cal C}(t)$, the negativity ${\cal N}(t)$, and the population
$\rho_{aa}(t)$ of the antisymmetric state for the initial state of both
atoms excited ($\rho_{ee}(0)=1$) and the interatomic distance
$r_{12}=\lambda/12$, which gives the collective damping
$\Gamma_{12}=0.95\,\Gamma$ and the  dipole-dipole interaction
frequency $\Omega_{12}=9.30\,\Gamma$. As expected, there is no
entanglement before the populations of the exited state and the
symmetric state depopulate, but some entanglement appears for
longer times, and the concurrence again becomes equal to the
population of the antisymmetric state. The values of the negativity in
this case are much smaller than the values of the concurrence, which
itself is very small. Exciting two atoms initially is thus very
ineffective in producing entanglement.

\begin{figure}[htb]
  \centering
  \includegraphics[height=9cm]{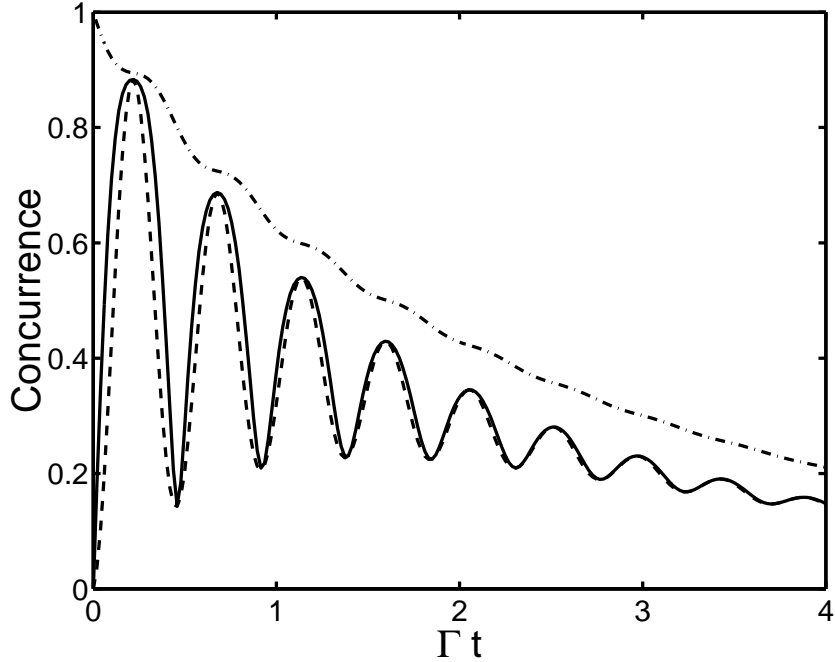}
  \caption{Time evolution of the concurrence ${\cal C}(t)$ (solid line),
  $\rho_{aa}(t)-\rho_{ss}(t)$ (dashed line), and
  $\rho_{aa}(t)+\rho_{ss}(t)$ (dashed-dotted line) for two
  nonidentical atoms with
  $\Delta=(\omega_{2}-\omega_{1})/2=10\,\Gamma$;  initially
atom 1 is excited ($\rho_{44}(0)=1$), $\mbox{\boldmath$\hat{\mu}$}
\perp\mbox{\boldmath$\hat{r}$}_{12}$, and $r_{12}=\lambda/12$
  ($\Gamma_{12}=0.95\,\Gamma$, $\Omega_{12}=9.30\,\Gamma$).}
  \label{fig:nonident}
\end{figure}

For nonidentical atoms, although the analytical solution is possible,
the formulas are rather lengthy and we will not adduce them
here. Instead, we plot in Fig.~\ref{fig:nonident} an example of the
evolution for the concurrence 
for the case of atom ``1'' excited ($\rho_{44}(0)=1$) with
$\Delta=(\omega_{2}-\omega_{1})/2=10\,\Gamma$ and $r_{12}=\lambda/12$
  ($\Gamma_{12}=0.95\,\Gamma$, $\Omega_{12}=9.30\,\Gamma$). This means
that we have $\Delta\sim\Omega_{12}$. As it is evident from the
equations of motion~(\ref{eq:eqsofmotion}),  for two nonidentical
atoms, {\em i.e.}, for $\Delta\ne 0$, there is a coupling between the
populations of the symmetric state and the asymmetric state, which
leads to a coherent transfer of population from one state to the
other. This introduces oscillations into the populations of both
states. Since the oscillations of the two populations, for the parameters of
Fig.~\ref{fig:nonident}, are  opposite in phase, they add up in
$\rho_{aa}(t)-\rho_{ss}(t)$ and subtract in $\rho_{aa}(t)+\rho_{ss}(t)$, as
clearly seen from the figure. The concurrence is oscillating,
similarly to the situation shown in Fig.~\ref{fig:conc1}, between the
lower bound ($\rho_{aa}(t)-\rho_{ss}(t)$) and the upper bound
$\rho_{aa}(t)+\rho_{ss}(t)$, but this time the lower bound itself
undergoes oscillations, which results in increasing the concurrence at
the maxima. The value at the first maximum is 0.88, which is higher
than the corresponding value for identical atoms equal to 0.86. It is
thus possible to enhance the transient entanglement in 
the two-atom system when the two atoms are nonidentical.

\section{Conclusion}
In this paper we have studied entanglement created in a system of two
two-level atoms via the spontaneous emission.  We have found
analytical formulas for the 
concurrence and the negativity, the two different measures of
entanglement usually used to quantify the amount of entanglement. 
Our formulas are valid for a broad class of initial conditions which
are represented by the block form of the density matrix.

We have shown that for short times, when initially only one atom is
excited, the amount of entanglement exhibits oscillatory behaviour
until the time at which the population of the symmetric states becomes zero.
For long times the concurrence becomes equal to the population
$\rho_{aa}(t)$ of the slowly decaying antisymmetric state.
For both atoms initially excited there are no oscillations, and the
entanglement appears only for long times, when only the antisymmetric
state contributes to the entanglement.

\end{document}